\documentclass[reprint, superscriptaddress, twocolumn, amsmath, amssymb, aps, prb]{revtex4-2}
\usepackage{graphicx}
\usepackage{dcolumn}
\usepackage{bm}
\usepackage{placeins}
\usepackage{appendix}
\usepackage{upgreek}

\usepackage{verbatim}
\usepackage[usenames,dvipsnames]{xcolor}
 \usepackage{amsmath}
 \usepackage{amsfonts}
 \usepackage{amssymb}
 \usepackage[colorlinks=true,citecolor=blue,linkcolor=red]{hyperref}

 \usepackage{bbold}     
 \usepackage[makeroom]{cancel}  
 \usepackage{multirow}    
 \usepackage[normalem]{ulem}        
 \usepackage{array}
 \usepackage{pdfpages}
\makeatletter
 \AtBeginDocument{\let\LS@rot\@undefined}
 \makeatother

\usepackage{lineno}

\begin{document}


\title{In situ tuning of dynamical Coulomb blockade on Andreev bound states in hybrid nanowire devices}

\author{Shan Zhang}
 \email{equal contribution}
\affiliation{State Key Laboratory of Low Dimensional Quantum Physics, Department of Physics, Tsinghua University, Beijing 100084, China}

\author{Zhichuan Wang}
\email{equal contribution}
\affiliation{Beijing National Laboratory for Condensed Matter Physics, Institute of Physics, Chinese Academy of Sciences, Beijing 100190, China}

\author{Dong Pan}
 \email{equal contribution}
\affiliation{State Key Laboratory of Superlattices and Microstructures, Institute of Semiconductors, Chinese Academy of Sciences, P. O. Box 912, Beijing 100083, China}

\author{Zhaoyu Wang}
\affiliation{State Key Laboratory of Low Dimensional Quantum Physics, Department of Physics, Tsinghua University, Beijing 100084, China}

\author{Zonglin Li}
\affiliation{State Key Laboratory of Low Dimensional Quantum Physics, Department of Physics, Tsinghua University, Beijing 100084, China}

\author{Zitong Zhang}
\affiliation{State Key Laboratory of Low Dimensional Quantum Physics, Department of Physics, Tsinghua University, Beijing 100084, China}

\author{Yichun Gao}
\affiliation{State Key Laboratory of Low Dimensional Quantum Physics, Department of Physics, Tsinghua University, Beijing 100084, China}

\author{Zhan Cao}
\affiliation{Beijing Academy of Quantum Information Sciences, Beijing 100193, China}

\author{Gu Zhang}
\affiliation{Beijing Academy of Quantum Information Sciences, Beijing 100193, China}

\author{Lei Liu}
\affiliation{State Key Laboratory of Superlattices and Microstructures, Institute of Semiconductors, Chinese Academy of Sciences, P. O. Box 912, Beijing 100083, China}

\author{Lianjun Wen}
\affiliation{State Key Laboratory of Superlattices and Microstructures, Institute of Semiconductors, Chinese Academy of Sciences, P. O. Box 912, Beijing 100083, China}

\author{Ran Zhuo}
\affiliation{State Key Laboratory of Superlattices and Microstructures, Institute of Semiconductors, Chinese Academy of Sciences, P. O. Box 912, Beijing 100083, China}

\author{Dong E. Liu}
\affiliation{State Key Laboratory of Low Dimensional Quantum Physics, Department of Physics, Tsinghua University, Beijing 100084, China}
\affiliation{Beijing Academy of Quantum Information Sciences, Beijing 100193, China}
\affiliation{Frontier Science Center for Quantum Information, Beijing 100084, China}

\author{Ke He}
\affiliation{State Key Laboratory of Low Dimensional Quantum Physics, Department of Physics, Tsinghua University, Beijing 100084, China}
\affiliation{Beijing Academy of Quantum Information Sciences, Beijing 100193, China}
\affiliation{Frontier Science Center for Quantum Information, Beijing 100084, China}

\author{Runan Shang}
 \email{shangrn@baqis.ac.cn}
\affiliation{Beijing Academy of Quantum Information Sciences, Beijing 100193, China}

\author{Jianhua Zhao}
 \email{jhzhao@semi.ac.cn}
\affiliation{State Key Laboratory of Superlattices and Microstructures, Institute of Semiconductors, Chinese Academy of Sciences, P. O. Box 912, Beijing 100083, China}

\author{Hao Zhang}
\email{hzquantum@mail.tsinghua.edu.cn}
\affiliation{State Key Laboratory of Low Dimensional Quantum Physics, Department of Physics, Tsinghua University, Beijing 100084, China}
\affiliation{Beijing Academy of Quantum Information Sciences, Beijing 100193, China}
\affiliation{Frontier Science Center for Quantum Information, Beijing 100084, China}


\begin{abstract}

Electron interactions in quantum devices can exhibit intriguing phenomena. One example is assembling an electronic device in series with an on-chip resistor. The quantum laws of electricity of the device is modified at low energies and temperatures by dissipative interactions induced by the resistor, a phenomenon known as dynamical Coulomb blockade (DCB). The DCB strength is usually non-adjustable in a fixed environment defined by the resistor. Here, we design an on-chip circuit for InAs-Al hybrid nanowires where the DCB strength can be gate-tuned in situ. InAs-Al nanowires could host Andreev or Majorana zero-energy states. This technique enables tracking the evolution of the same state while tuning the DCB strength from weak to strong. We observe the transition from a zero-bias conductance peak to split peaks for Andreev zero-energy states. Our technique opens the door to in situ tuning interaction strength on zero-energy states.

\end{abstract}

\maketitle

Hybrid semiconductor-superconductor nanowires provide an excellent platform for the realization of various quantum electronic devices such as supercurrent transistors \cite{Leo_Supercurrent}, Cooper pair splitters \cite{Nature_Cooper_pair}, gate-tunable qubits \cite{2015_PRL_gatemon}, Andreev quantum point contacts \cite{Zhang2017Ballistic} and possible Majorana zero modes (MZMs) \cite{Prada2020,NextSteps,Review_Marra}.  These devices can combine the pairing correlations from the superconductor with low dimensional gate-tunable carrier densities from the semiconductor, thus attracting much interest in both fundamental and application-wise research perspectives. A fascinating prediction is the MZMs \cite{Lutchyn2010, Oreg2010} which hold promise toward topological quantum computations. Though enormous experimental efforts have been spent in the quest for MZM signatures \cite{Mourik, Deng2016, Gul2018, Zhang2021, Song2022, WangZhaoyu}, the major uncertainty comes from Andreev bound states (ABSs) \cite{Andreev1966} which can form in the same device and mimic MZMs \cite{Prada2012, BrouwerSmooth, Silvano2014, Liu2017, Loss2018ABS, TudorQuasi, WimmerQuasi, GoodBadUgly}. Both MZMs and ABSs can lead to similar zero-bias peaks (ZBPs) in tunneling conductance spectroscopy. Therefore, one of the top priorities along this roadmap is finding an effective diagnostic tool to distinguish these two scenarios. 

Interactions could stabilize MZMs and suppress ABS signals. The mechanism is that interaction-induced renormalization sharpens the transitions between different physics to different fixed points \cite{cardy_1996}. One example is by adding an on-chip resistor (a few k$\Omega$) in series with the MZM device \cite{Dong_PRL2013}. The resistor provides a dissipative electromagnetic environment. Small tunnel junctions embedded in this interactive environment can reveal a conductance dip near zero energy at low temperatures \cite{Delsing_1989, Devoret_1990, Ingold, Flensberg_1992, Pierre_2001_PRL, Pierre_2007_PRL, Pierre2011, Gleb_Nature, Gleb_NaturePhysics, DCB_PRL_2013, Dong_PRB2014, Jezouin_2013,Pierre_PRX, Flensberg_ECB, DCB_PRL_2020}. This ``dissipative tunneling'' is also dubbed environmental Coulomb blockade or dynamical Coulomb blockade (DCB). Since ``dissipation'' in the hybrid nanowire literatures usually refers to soft gaps \cite{Takei2013, LiuDissipation}, to avoid confusion, in this paper we adopt the term ``DCB'' to describe this interaction effect. The DCB strength scales with the environmental impedance that is determined by the on-chip resistor. A larger resistor leads to a stronger suppression of the zero-bias conductance. The key idea of the proposal \cite{Dong_PRL2013, Dong_2021} is that increasing the DCB strength splits ABS-induced ZBPs by suppressing the conductance at zero bias. As for MZMs, the ZBP is mediated by symmetric resonant Andreev reflections \cite{DasSarma2001, Law2009, Flensberg2010, Tanaka}. The topological ZBP could survive the DCB suppression and stay at the quantized value of $2e^2/h$ as long as the DCB strength is below a threshold (environmental impedance less than the resistance quantum $h/2e^2$) \cite{Dong_PRL2013}.

\begin{figure}[b]
\includegraphics[width=\columnwidth]{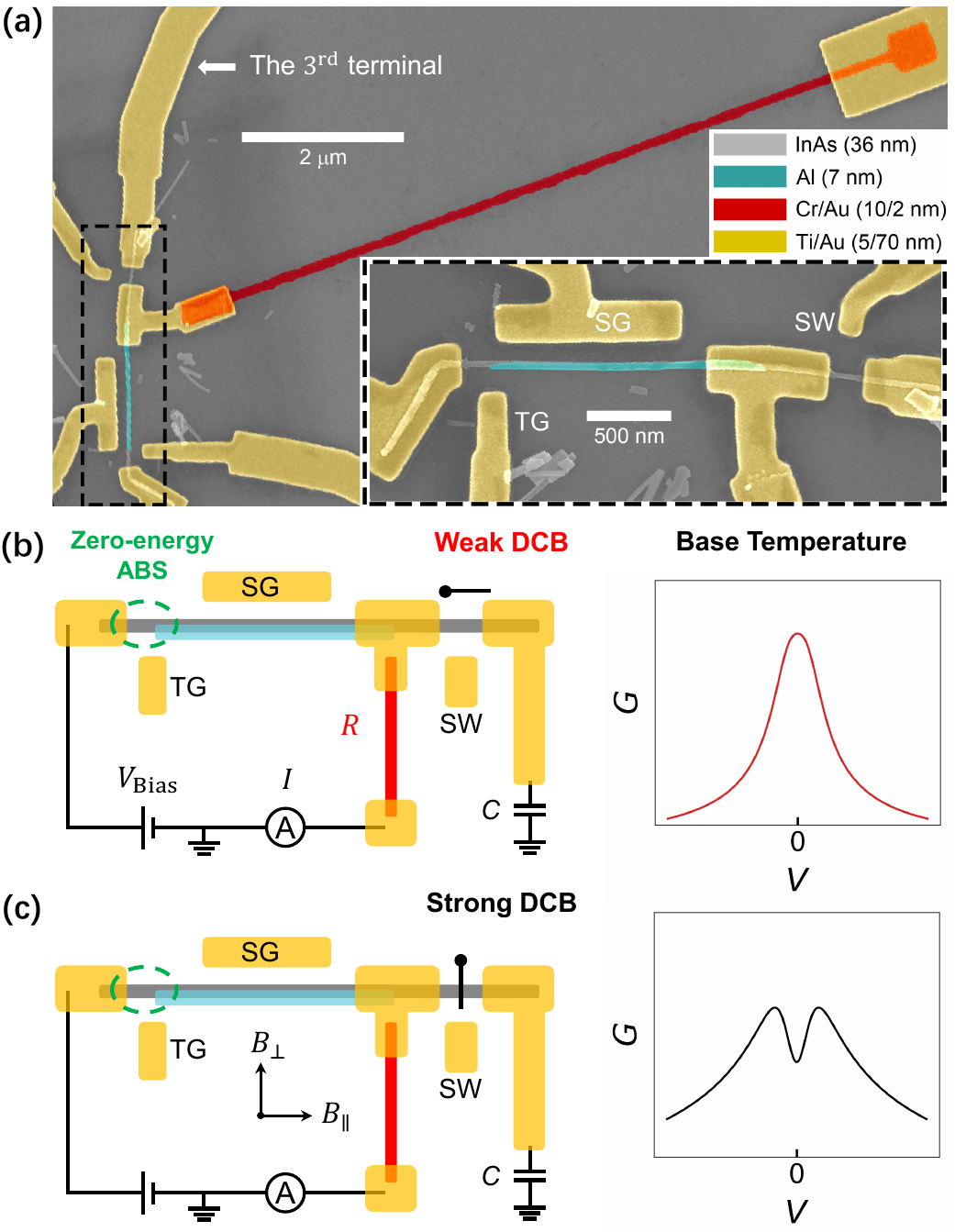}
\centering
\caption{(a) SEM of device A (false-colored). Inset, zoom-in of the InAs-Al part. The substrate is $p$-doped Si covered by 300 nm thick SiO$_2$. The Si (back gate) was grounded through out the measurement. (b) Left, a simplified schematic of the circuit diagram. The switch (SW) gate turns on the InAs segment nearby, shunting the dissipative resistor (weak DCB). Right, a zero-energy ABS resolves a ZBP (not in scale) at a finite fridge temperature. (c) Left, SW pinches off the InAs segment (strong DCB). Right, the same zero-energy ABS resolves split peaks due to the suppression of $G$ at zero bias. }
\label{fig1}
\end{figure}

In recent experiments \cite{ZhangShan, WangZhichuan}, we have introduced DCB into hybrid InAs-Al nanowire devices where the ABS-induced ZBPs can be indeed suppressed, leading to split peaks. The circuit design is two-terminal where the nanowire and the resistor are connected in series. The drawback of this design is that the DCB strength is non-adjustable for a fixed barrier once the resistor has been fabricated. This makes it difficult to judge whether the measured split peaks originate from a zero-energy ABS (an effect of DCB) or just a finite-energy ABS, which can result in split peaks even without DCB. Moreover, the resistance of the resistor could only be estimated indirectly, introducing another error source. This possible error can also affect the accuracy of ZBP heights since the nanowire resistance is obtained by subtracting the resistor resistance from the total resistance of the two-terminal circuit. In this paper, we demonstrate a new circuit with a three-terminal design that solves these problems. The DCB strength can be in situ tuned using a gate voltage on the third terminal. This technique enables us to increase the DCB strength and track the evolution of the same zero-energy ABS, i.e. a ZBP gradually evolves to split peaks . The resistances of the resistor and the nanowire can also be accurately measured.  Our technique can not only serve as a powerful knob in the future Majorana detection, but also can facilitate more interaction-related physics experiments, e.g. the realization of Berezinsky-Kosterlitz-Thouless quantum phase transition or emulating two-channel Kondo physics \cite{Dong_PRB2014}. 

Figure 1(a) shows the scanning electron micrograph (SEM) of device A. An InAs-Al wire is contacted by three Ti/Au electrodes (yellow). The middle electrode is connected in series with a thin resistive Cr/Au film (red, 10 nm/2 nm in thickness), serving as the on-chip resistor. The resistance of the film, $R \sim$ 11.7 k$\Omega$, can be directly measured using the three-terminal set-up, see Fig. S1 in the supplementary material (SM) for details \cite{SM}. Directly measuring $R$ was not feasible in previous two-terminal designs, where the value of $R$ can only be indirectly inferred \cite{ZhangShan, WangZhichuan}. The wire growth details can be found at Ref. \cite{PanCPL}. Three side gates (yellow, Ti/Au, thickness 5 nm/70 nm) are used: TG tunes the barrier part; SG tunes the proximitized bulk; SW serves as a DCB switch. The middle electrode forms ohmic contact with the nanowire through Ar plasma etching. In this way, the nanowire is separated into two quantum conductors which are connected by the middle electrode. And the two segments of the nanowire (the barrier region and the SW-region) can be treated separately, avoiding uncontrolled non-local effects.

The basic idea is sketched in Figs. 1(b) and 1(c). A voltage, $V_{\text{Bias}}$, is applied to the first electrode. The current $I$ is measured after passing through the InAs-Al wire, the second electrode and the film ($R$). The differential conductance of the InAs-Al wire is calculated as $G=dI/dV$, where $V = V_{\text{Bias}}-I\times R$. The remaining InAs segment is connected to the third electrode (terminal), which extends all the way to the bonding pad on the chip. The bonding pad is connected to a fridge line whose room-temperature end is floated. The large capacitance of the pad ($C \sim$ 10 pF) can short-circuit the high-frequency quantum fluctuations. Since the DCB strength is determined by the environmental impedance regarding a frequency range from 0 to tens of GHz, the capacitor $C$ can significantly reduce this high-frequency impedance if the InAs segment is opened by SW. In this way, even though the dc current solely flows through the InAs-Al wire and $R$, the third terminal (opened for dc but grounded for high frequency) can reduce the DCB effect induced by $R$. We call this regime the weak DCB. A zero-energy ABS, formed near the barrier, will be resolved as a ZBP at low (but finite) temperatures. If  0 K could be reached, no matter how weak the DCB strength is, the zero-bias tunneling $G$ could always be suppressed to zero, splitting the ABS-induced ZBPs. More negative gate voltage on SW can pinch off the InAs segment and restore the DCB induced by $R$ (Fig. 1(c)). In this strong DCB regime, an ABS-induced ZBP will split due to the strong suppression of $G$ at zero bias. The right panels sketch the conductance line shapes (not in scale) in the two cases. We note that similar ideas of tuning DCB using a gate switch and large capacitors, i.e. the three-terminal geometry, have been previous implemented in the system of two dimensional electron gases \cite{Pierre_2007_PRL, Pierre2011,Jezouin_2013}.

\begin{figure}[htb]
\includegraphics[width=\columnwidth]{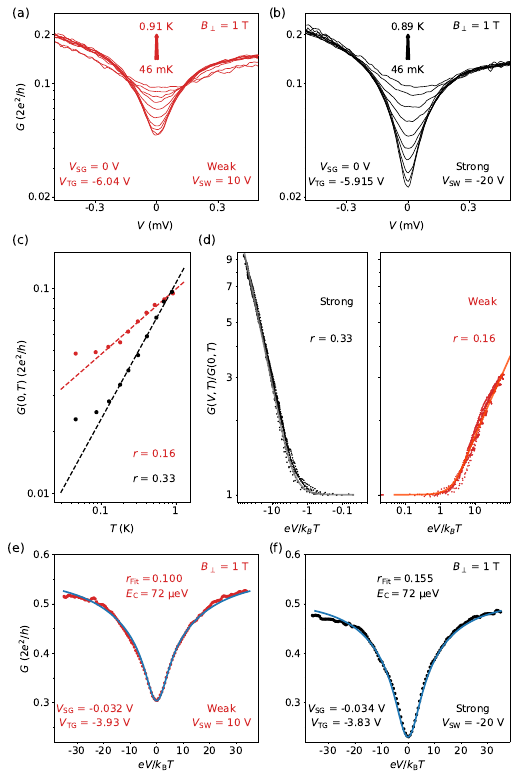}
\centering
\caption{(a) Zero-bias suppression of the normal state $G$ ($B_{\perp}$ = 1 T) in the weak DCB regime ($V_{\text{SW}}$ = 10 V). $V_{\text{TG}}$ = -6.04 V, $V_{\text{SG}}$ = 0 V. The fridge $T$ varies from 46 mK to 0.91 K for different curves. (b) $T$ dependence in the strong DCB regime ($V_{\text{SW}}$ = -20 V) at a similar barrier transmission. $V_{\text{TG}}$ = -5.915 V, $V_{\text{SG}}$ = 0 V. (c) $T$ dependence of the zero-bias $G$, extracted from (a) (red) and (b) (black). Dashed lines are the power law fits with the exponents ($r$) labeled. (d) Replotting all the curves in (a) (right) and (b) (left) using dimensionless units. The solid lines are fits using the extracted $r$'s from (c). (e) The DCB dip at base $T$ in the weak DCB regime with a higher barrier-transmission. The blue solid line is the fit using the $P(E)$ theory. The effective temperature is assumed to be $\sim$100 mK. The charging energy is extracted to be 72 $\upmu$eV. (f) Similar to (e) but in the strong DCB regime.   }
\label{fig2}
\end{figure}

In Figure 2, we characterize the DCB strength in device A. A perpendicular magnetic field of 1 T (see Fig. 1(c) for the $B$ orientation) drives the device normal. DCB effects in a normal tunnel junction have been well studied \cite{Delsing_1989, Devoret_1990, Ingold, Flensberg_1992}. Figures 2(a) and 2(b) show the characteristic suppression near zero-bias and the $T$ dependence at $V_{\text{SW}}$ = 10 V (segment fully opened, weak DCB) and -20 V (segment pinched off, strong DCB). For the $V_{\text{SW}}$ dependence of the segment conductance, see Fig. S1 in SM \cite{SM}. The suppression of zero-bias $G$ (at base $T$) in the weak DCB regime is indeed weaker than the strong DCB case. Note that the $y$ axis is in the logarithmic scale to highlight this difference. The tunnel transmissions for Figs. 2(a) and 2(b) are similar, reflected by the $G$'s at high $T$ (0.9 K). Keeping the same transmission is important for this comparison since the DCB strength can also be modified by the barrier transparency and scales with the Fano factor \cite{Pierre_2007_PRL, Pierre2011, DCB_PRL_2020}.

A more quantitative description of the ``zero-bias dip'' is the power law: $G\propto \text{max}(k_{\text{B}}T, eV)^{2r}$ \cite{Ingold}. The exponent $r$ determines the effective environmental impedance ($r \times h/e^2$) and the DCB strength. Figure 2(c) shows the extracted zero-bias $G$'s from Figs. 2(a) and 2(b). The dashed lines are the temperature power-law fits ($G \propto T^{2r}$), revealing $r$ of 0.16 for the weak and 0.33 for the strong DCB cases. The deviations below 100 mK suggests a gradual saturation of the device electron $T$. 

Figure 2(d) plots all the curves in Figs. 2(a) and 2(b) (over half of the bias branch) using the dimensionless units: $G(V, T)/G(0, T)$ for the $y$ axis and $eV/k_{\text{B}}T$ for the $x$ axis. All curves ``collapse'' onto a single universal line with minor deviations (the grey line for the strong and the red for the weak DCB regimes). The ``linear trend'' in this log-log plot for the regions of $|eV/k_{\text{B}}T|>1$ suggests the power law for $V$. The universal line is a prediction of the conventional DCB theory and obtained by performing numerical differentiation on the formula \cite{Gleb_Nature}:   $I(V,T) \propto VT^{2r}|\Gamma (r+1+ieV/2\pi k_{\text{B}}T)/\Gamma (1+ieV/2\pi k_{\text{B}}T)|^2$, where $\Gamma$ is the Gamma function. $r$ is extracted from Fig. 2(c). For $T<$ 100 mK, we used the electron $T$ (extracted from Fig. 2(c)) for the $x$ axis in Fig. 2(d). For more power-law analysis and its SW tunability, see Fig. S2 for device A, Fig. S3 for device B and Fig. S4 for device C in SM \cite{SM}.

\begin{figure}[htb]
\includegraphics[width=\columnwidth]{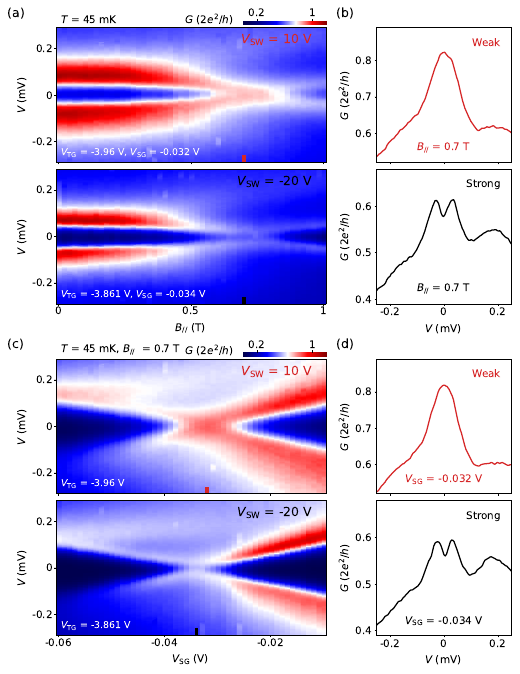}
\centering
\caption{(a) $B$ dependence of an ABS in the weak (upper, $V_{\text{SW}}$ = 10 V) and strong (lower, $V_{\text{SW}}$ = -20 V) DCB regimes. (b) Line cuts at $B =$ 0.7 T. (c) Gate dependence of this ABS at 0.7 T in the weak (upper) and strong (lower) DCB regimes. (d) Line cuts of (c) at the ``crossing points''.  }
\label{fig3}
\end{figure}

The power-law holds over only a half order of magnitude (Fig. 2(c)), due to a relatively small dissipative resistor $R$. In our previous work \cite{ZhangShan} where $R$ is much larger, the power-law range can exceed an order of magnitude.  Since a power-law over a short range has a limited implication, to better characterize the DCB dip we have performed the $P(E)$ calculation \cite{Ingold, Jezouin_2013, Flensberg_1991}. $P(E)$ theory can capture the conductance features at higher energies (outside the cut-off energy where the power-law no longer holds). Figures 2(e) and 2(f) shows the $P(E)$ fitting (blue curves) in the weak and strong DCB regimes, agreeing reasonable well with the experiment. The input parameters are the charging energy of the tunnel junction ($E_{\text{C}}$ = 72 $\upmu$eV), the effective $T$ ($\sim$ 100 mK), and $r$ as  the fitting parameter, see SM for the calculation details. The fitted $r$ ($r_{\text{Fit}}$) is indeed smaller in the weak DCB regime than that in the strong DCB regime.

After establishing the in situ tunability of DCB strength in the normal state regime, we now study ABSs in the superconducting regime by aligning $B$ parallel to the nanowire axis (device A). In Fig. 3(a), a finite-energy ABS at zero field moves to zero energy at $B \sim$ 0.7 T. In the weak DCB regime (upper), this process reveals a level crossing and a ZBP at 0.7 T (Fig. 3(b)). In the strong DCB regime (lower), the crossing becomes anti-crossing and the ZBP splits at 0.7 T. This peak splitting does not originate from a finite-energy ABS but reflects the DCB effect on a zero-energy ABS. We can confirm this by tracing the same state back to the weak DCB regime. The small differences of $V_{\text{TG}}$ (and $V_{\text{SG}}$) between the weak and strong DCB regimes are to compensate for the residue crosstalk with SW, ensuring that the same ABS is being monitored during the DCB tuning.

\begin{figure}[b]
\includegraphics[width=\columnwidth]{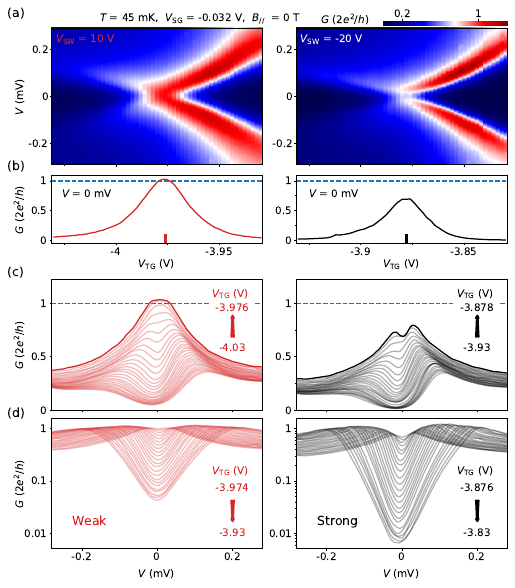}
\centering
\caption{(a) Gate dependence of a ZBP whose height is fine-tuned to $2e^2/h$ in the weak (left) and strong (right) DCB regimes. (b) Zero-bias line cuts from (a). (c) Waterfall plots for the left half of the gate ranges in (a). The top curves correspond the gate voltages indicated by the bars in (b). (d) Waterfall plots for the other half of the gate range. The $y$ axis is in logarithmic scale. }
\label{fig4}
\end{figure}

The transition from ``level-crossing-ZBP'' in the weak DCB regime to ``anti-crossing-split-peaks'' in the strong DCB regime for the same ABS is enabled by our new technique: the three-terminal circuit design (the main advancement of this work). This transition can also be revealed in the gate scans of the same ABS, see Figs. 3(c) and 3(d). The overall $G$ of the ABS in Fig. 3, $\sim$$e^2/h$, is significantly higher than that in Fig. 2, suggesting a higher barrier transparency. Higher transparency can ``weaken'' the DCB effect \cite{Pierre_2007_PRL, Pierre2011, DCB_PRL_2020}, see also Figs. 2(a-b) vs 2(e-f). To verify the validity of our technique in this high transparency regime, in Fig. S5 in SM \cite{SM} we show the power law of the normal state $G$ and the SW-tunable $r$ at $G \sim$$e^2/h$. Note that the suppression of conductance in the normal state and superconducting (ABS) regime may differ. In the normal state, $G \propto T^{2r}$, while for ABS, $G \propto T^{4r}$ or $T^{8r}$ due to Andreev reflections.  The exponents of $8r$ and $4r$ originate from the coherent and incoherent Andreev reflections, respectively \cite{Dong_2021}. Other values of exponents are also possible when different processes mix together. Nevertheless, the tunable $r$ (by SW) combined with the qualitative change from ZBP to split peaks constitute a powerful tool to in situ identify ABSs. Figure S6 in SM \cite{SM} shows additional scans for intermediate $V_{\text{SW}}$, ranging from 10 V and -20 V, as well as $V_{\text{TG}}$ scans.

Whether tuning the DCB strength could split an ABS-induced ZBP or not depends on many factors, e.g. the device temperature, the details of the ABS and to what extend can $r$ be tuned or varied.  Figures S7 and S8 in SM \cite{SM} show results from device B where the tunable range of $r$ is smaller: from $\sim$0.05 to 0.13. As a result, the suppression (splitting) of ABS-induced ZBPs is less effective: only part of the ZBP regions in the parameter space show the splitting by increasing $r$ (Fig. S7). Figure S8 monitors the continuous change of a finite-energy ABS.

The ZBPs so far are smaller than $2e^2/h$, obviously not MZMs but ABS-induced. A natural follow-up question is: how will $2e^2/h$-ZBPs evolve by varying the DCB strength? A previous theoretical work \cite{Dong_PRL2013} predicts that the height of MZM-induced ZBPs will not be affected for $r <0.5$ and will robustly stick to the quantized value. While ZBPs forming $2e^2/h$ plateaus are rare \cite{WangZhaoyu}, finding them and testing their stability by tuning $r$ are the goal of our future study. In Figure 4 we address a related question by studying a ZBP fine-tuned to $2e^2/h$. 

Figure 4(a) (the left panel) shows the gate dependence of this ABS in device A. MZMs are not expected since $B$ = 0 T. The ZBP height is accidentally close to $2e^2/h$. This $2e^2/h$-peak is fine-tuned since the zero-bias $G$ does not resolve a plateau (Fig. 4(b), the red curve). Increasing the DCB strength from weak to strong, the zero-bias $G$ gets significantly suppressed (Fig. 4(b), the black curve). In addition, the ZBP in the weak DCB regime becomes split peaks in the strong DCB regime, see the top red and black line cuts in Fig. 4(c) and the color-maps in Fig. 4(a). The zero-bias $G$ of split peaks in the weak DCB regime, corresponding finite-energy ABSs, is also suppressed when tuned into the strong DCB regime. In Fig. 4(c) we show this suppression in linear scale to highlight the high $G$ part. The $V_{\text{TG}}$ ranges correspond to the left sides of the bars in Fig. 4(b). In Fig. 4(d), logarithmic scale is used to highlight the difference in the lower $G$ regimes.  The $V_{\text{TG}}$ ranges correspond to the right sides of the bars in Fig. 4(b). For the $V_{\text{SG}}$ dependence of this ABS, see Fig. S9 in SM \cite{SM}.

ZBPs fine-tuned to $2e^2/h$ have recently raised intense discussion on possible false-positive signatures of MZMs \cite{GoodBadUgly}. Figure 4 shows that these peaks could be identified and safely ``filtered out'' by our technique in the search of true quantized peaks.

To summarize, we have established a three-terminal circuit technique which enables the in situ tuning of the DCB strength in hybrid semiconductor-superconductor nanowire devices. This technique allows to ``diagnose'' the origins of zero-energy states by changing the interaction strength. Zero-bias peaks induced by Andreev bound states with heights at non-quantized values or accidentally at $2e^2/h$ can be split (filtered out) by increasing the DCB strength. The tunability of the DCB strength is determined by the SW segment of the nanowire. Using a second (thick) nanowire as the SW segment could improve this tunability.  By tuning interaction in situ, our result paves the way toward finding cleaner and stronger evidence of Majorana zero modes in future studies.

\section{Acknowledgment} 

This work is supported by Tsinghua University Initiative Scientific Research Program, National Natural Science Foundation of China (Grant Nos. 12104053, 92065106, 92065206, 61974138, 12004040, 11974198, 12374158). D.P. also acknowledges the support from Youth Innovation Promotion Association, Chinese Academy of Sciences (Nos. 2017156 and Y2021043). Raw data and processing codes within this paper are available at https://doi.org/10.5281/zenodo.7317835

\bibliography{mybibfile}

\newpage

\onecolumngrid

\newpage


\section*{Supplementary material}

# Supplementary Information for "In situ tuning of dynamical Coulomb blockade on Andreev bound states in hybrid nanowire devices"

Shan Zhang,[1,][*] Zhichuan Wang,[2,][*] Dong Pan,[3,][*] Zhaoyu Wang,[1] Zonglin Li,[1] Zitong Zhang,[1] Yichun Gao,[1] Zhan Cao,[4] Gu Zhang,[4] Lei Liu,[3] Lianjun Wen,[3] Ran Zhuo,[3] Dong E. Liu,[1, 4, 5] Ke He,[1, 4, 5] Runan Shang,[4, †] Jianhua Zhao,[3, ‡] and Hao Zhang[1, 4, 5, §]

[1]*State Key Laboratory of Low Dimensional Quantum Physics,*
*Department of Physics, Tsinghua University, Beijing 100084, China*
[2]*Beijing National Laboratory for Condensed Matter Physics,*
*Institute of Physics, Chinese Academy of Sciences, Beijing 100190, China*
[3]*State Key Laboratory of Superlattices and Microstructures, Institute of Semiconductors,*
*Chinese Academy of Sciences, P. O. Box 912, Beijing 100083, China*
[4]*Beijing Academy of Quantum Information Sciences, Beijing 100193, China*
[5]*Frontier Science Center for Quantum Information, Beijing 100084, China*

**Device Fabrication.** The InAs-Al nanowires were transferred from the growth chip via clean room tissues onto a highly p-doped Si chip covered with 300-nm-thick silicon dioxides. Selective etching of Al on InAs-Al nanowires was performed for 10 seconds at 50 °C using Transene Aluminum Etchant Type D. The etch window was patterned using electron beam lithography (EBL). Then, Cr/Au films (10 nm/2 nm in thickness) were evaporated next to the nanowires, acting as the dissipative resistors. Ti/Au contacts and gates were evaporated in another EBL round. Before the evaporation, argon plasma etching was performed for 70 s at a power of 55 W and a pressure of 52 mTorr to obtain good ohmic contacts.

**Measurement Setup.** All transport measurements were performed in a Bluefors dilution refrigerator equipped with a 6-1-1 T vector magnet, except for the measurements of Figs. S1(c)-(f) which were carried out in an Oxford dilution fridge in another cool down. Copper powder filters and RC filters were placed on the mixing chamber plates of both dilution refrigerators. The differential conductance was measured at 40.17 Hz using a standard lock-in technique. The detailed measurement parameters for each data set can be found in the corresponding raw data file.

**Environmental Impedance.** The environmental impedance $R_\text{E}$ sensed by an ABS is determined by both the on-chip resistor $R$ and the third terminal, i.e. the SW segment of the nanowire $R_\text{SW}$, as $1/R_\text{E} = 1/R + 1/R_\text{SW}$. The resistance of the Cr/Au thin film (on-chip resistor, $R$) is 11.76 k$\Omega$ (see Fig. S1(f)), corresponding to $r = 0.46$. When the SW segment of the nanowire is pinched off, $R_\text{SW} \gg R$. $R_\text{E} \sim R$. The fitted dissipation strength $r \sim 0.33$ (extracted from the power-law) is slightly smaller than 0.46. This deviation is because that the formula above applies only when $R_\text{E}$ is much smaller than the tunneling resistance ($R_\text{T}$) of the tunnel junction (the normal state resistance in Fig. 2). When $R_\text{E}$ is comparable to ($R_\text{T}$), the effective impedance should then be replaced by $1/(1/R_\text{E}+1/R_\text{T})$ [1]. Taking this effect into account, we can get $r = 0.38$, roughly consistent with the fitted value. When the SW segment is fully opened ($R_\text{SW} \sim 8$ k$\Omega$), $r$ is calculated to be 0.18, roughly consistent with the fitted r $\sim$ 0.16.

**Tunability of the dissipation strength.** The tunability of $r$ for the devices in this paper is limited by both $R$ and $R_\text{SW}$. To enhance this tunability for future devices, $R$ can be designed larger than $h/2e^2$ so that the maximum $r$ can exceed 0.5. The Cr/Au thin film can be designed longer to fulfill this goal. Tuning $r$ to a small value (close to 0) requires that $R_\text{SW}$ can be gate-tuned from infinite (pinched off) to low resistance (opened). Currently, this low resistance value is 8 k$\Omega$, possibly due to the thin diameter of the wire. In the future, a thick InAs nanowire can be used as the switch segment to fulfill this goal.

**$P(E)$ Calculations.** We adopt the $P(E)$ theory [2] to fit the DCB curves in Figs. 2(e-f): $P(E) = \exp[J(t)+iEt/\hbar]$, where the equilibrium phase correlation function $J(t) =< [\phi(t) − \phi(0)], \phi(0) >$. The environmental impedance $Z_\text{t}(\omega) = 1/(iRC\omega + 1)$, where $R = r \times h/e^2$ is the dissipative resistance. $r$ is assumed to be 0.1 and 0.155 for Figs. 2(e) and 2(f), respectively. $C$ is the tunnel junction capacitance, estimated based on the charging energy (72 µeV, see Fig. S5(d)). $\omega/2\pi$ is the frequency. $T$ is around 100 mK based on Fig. 2(c). Following the calculation in Ref.[3], $J(t) = \pi \frac{R}{h/e^2}[(1-e^{-|t|/RC})(\cot\frac{\hbar}{2RCk_\text{B}T}-i) - \frac{2k_\text{B}T|t|}{\hbar} + 2\sum_{n=1}^{+\infty}\frac{1-e^{-\omega_n|t|}}{2\pi n[(RC\omega_n)^2-1]}]$. $\omega_n = 2\pi n k_\text{B}T/\hbar$ is the Matsubara frequency, and $n$ is an integer. The differential conductance $G(V,T) \equiv dI/dV = \frac{1}{R_\text{T}}[1+2\int_0^{+\infty} \pi t(\frac{k_\text{B}T}{\hbar})^2 \times \text{Im}[e^{J(t)}]\cos\frac{eVt}{\hbar}\sinh^{-2}\frac{\pi t k_\text{B}T}{\hbar}dt]$ can then be calculated.

———

[*] equal contribution
[†] shangrn@baqis.ac.cn
[‡] jhzhao@semi.ac.cn
[§] hzquantum@mail.tsinghua.edu.cn

FIG. S1. Circuit basic calibration. (a) Simplified schematic of the circuit diagram for the switch (SW) channel calibration. (b) The channel conductance ($G_{SW}$) as a function of SW gate voltage. The filter resistance and $R$ are subtracted: $G = 1/(dV_{Bias}/dI - R - 2 \times R_{Filter})$. The bias voltage is set to 1 mV to reduce possible effects of superconductivity and dynamic Coulomb blockade (DCB). The SW channel is pinched off (for the black and gray curves) at $V_{SW} = -20$V, corresponding to the strong DCB regime. The weak DCB regime corresponds to the SW channel being opened ($V_{SW} = 10$V). The black and gray curves were measured closely before the measurement of Figs. 2-4 while the red and rose curves were measured after. (c) Four-terminal circuit for the dissipative film calibration. (d) Differential resistance of the film, $R_{Film} = dV/dI$, as a function of bias voltage and magnetic field ($B$). (e) I-V curves of the dissiaptive film at different $B$ values. The slope resolves a resistance of 11.76 k$\Omega$, slightly larger than the value in (d) (the lock-in method). (f) Temperature dependence of the film resistance using lock-in and dc methods. Resistance from the lock-in method is ~200 $\Omega$ lower than the dc values. This is because that $G_{SW}$ was low (corresponds to ~ 46 k$\Omega$) in (d). Together with the lead and junction capacitance (~15 nF) they act as a low-pass filter for the lock-in (measurement frequency 40.13 Hz). The dc value is more accurate and was used for $R_{Film}$ throughout the analysis. The film resistance shows little change over $B$, bias and temperature, therefore could be treated as a constant value. (g) Three-terminal circuit for calibrating the dissipative films of device B and device C. (h) The film and filter resistance, $R_{Filter} + R_{Film} = dV/dI$, in device B as a function of bias and $B$. (i) I-V curves for device B whose slope matches the values in (h). Note that $G_{SW}$ here is high (corresponds to ~ 3 k$\Omega$), therefore the dc and lock-in methods gave the same resistance value. (j) Temperature dependence of the dissipative film in device B. The filter resistance is subtracted. The error bars are estimated based on the fluctuations in (h). (k) I-V curve of the dissipative film (and one fridge filter) in device C at 140 mK. Black dots are measured data and the line is the linear fit.

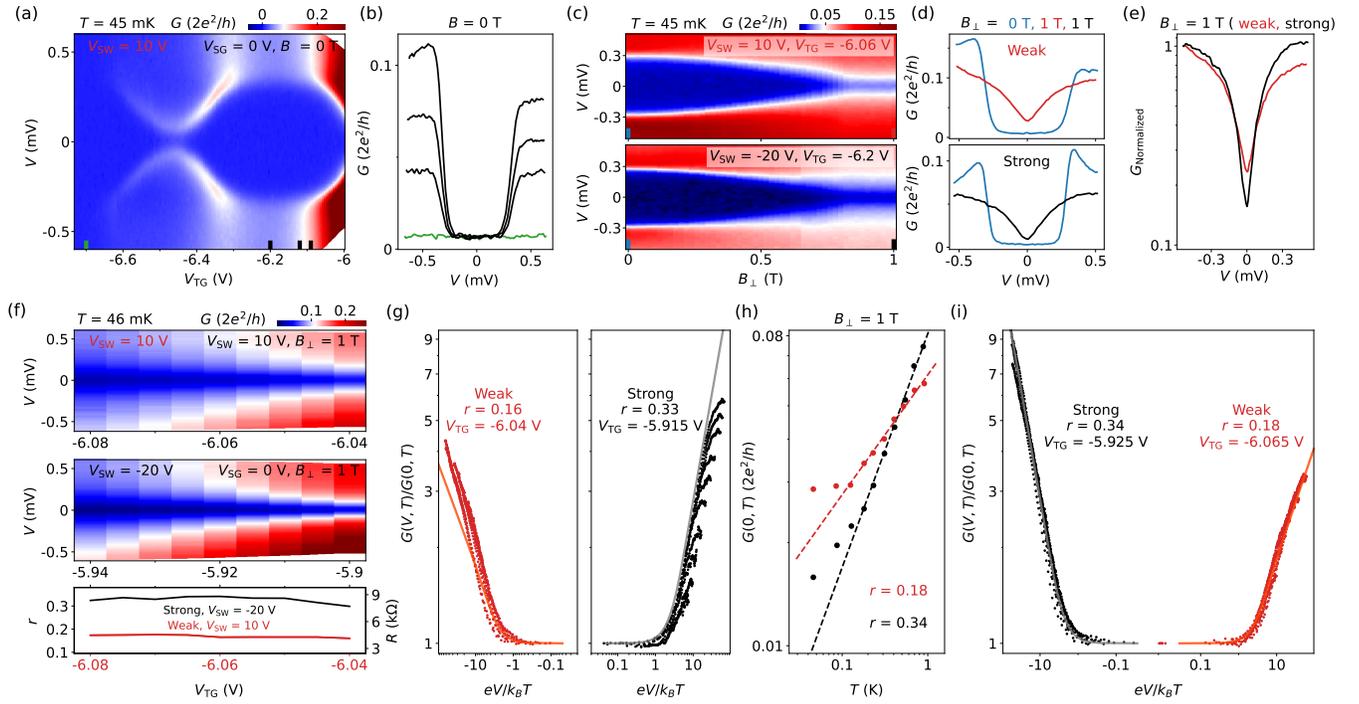

FIG. S2. DCB characterization in device A. (a) Superconducting gap calibration in the weak DCB regime ($V_{SW} = 10$ V) with line cuts shown in (b). The green curve is in the pinched off regime with a relatively large residue conductance background. (c) Gap-DCB transition driven by a perpendicular $B$ (in-plane) in the weak (upper) and strong (lower) DCB regimes. (d) Line cuts of (c) at 0 T (blue) and 1 T, resolving the gap and DCB dip, respectively. (e) Re-plotting the red and black curves in (d) in the logarithmic scale to highlight the low conductance difference. Note that the conductance is normalized by its value at the most negative bias. The DCB suppression (dip) for the black curve is stronger than for the red curve. (f) Gate dependence of the normal state DCB dip in the weak (upper) and strong (middle) DCB regimes. The lower panel illustrates the power-law exponent $r$, extracted from the temperature dependence of the zero-bias conductance ($T^{2r}$). The right axis translates this $r$ to the effective impedance $R = r \times h/e^2$. The strong DCB regime shows a larger $r$ value than that in the weak DCB regime. The data and analysis in Fig. 2 correspond to one line cut in (f). (g) Plotting the other half bias range of Figs. 2(a) and 2(b) using dimensionless unit (like Fig. 2(d)). (h) Temperature dependence of the zero-bias conductance at a different $V_{TG}$ value (another line cut from (f)) in the weak (red) and strong (black) DCB regimes. (i) Plotting all the curves (at different temperatures and half of the bias voltage range) at this gate voltage using dimensionless units, similar to Fig. 2(d).

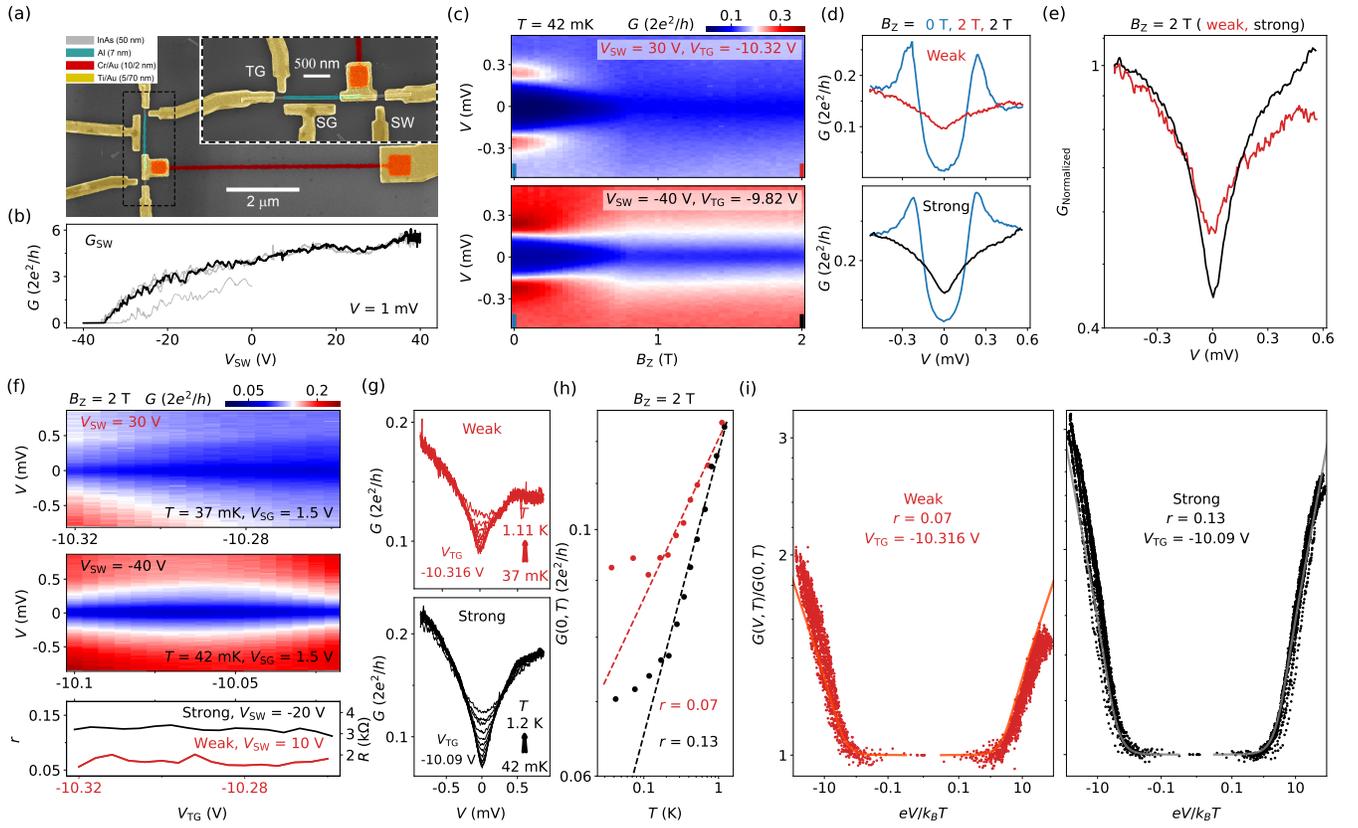

FIG. S3. DCB characterization in device B. (a) Device SEM. (b) Conductance of the SW channel as a function of $V_{SW}$. The bias voltage is 1 mV. The black and gray curves correspond to different measurements. Based on these pinch-off curves, the weak DCB regime corresponds to $V_{SW} = 30$ V while for the strong DCB regime $V_{SW} = -40$ V. (c-f) DCB calibration of device B in the weak and strong regimes, similar to Figs. S2(c)-2(f). (g) Temperature dependence of the DCB dip in the weak (upper) and strong (lower) regimes, corresponding to a line cut in (f). (h) The zero-bias conductance extracted from (g) in the weak and strong DCB regimes. The power-law exponent $r$ was extracted by fitting the range of $T > 100$ mK. (i) Re-plotting (g) using dimensionless units. The $x$ axis has a linear scale from -0.1 to 0.1 and a log scale outside. The lines are the universal curves, obtained based on $I(V, T) \propto V T^{2r} |\Gamma\left(r + 1 + ieV/2\pi k_B T\right)/\Gamma\left(1 + ieV/2\pi k_B T\right)|^2$, where $\Gamma$ is the Gamma function, $r$ is the power-law exponent extracted from (h), and $T$ is the temperature. For $T < 100$ mK, the electron temperature extracted from the power law in (h) is used.

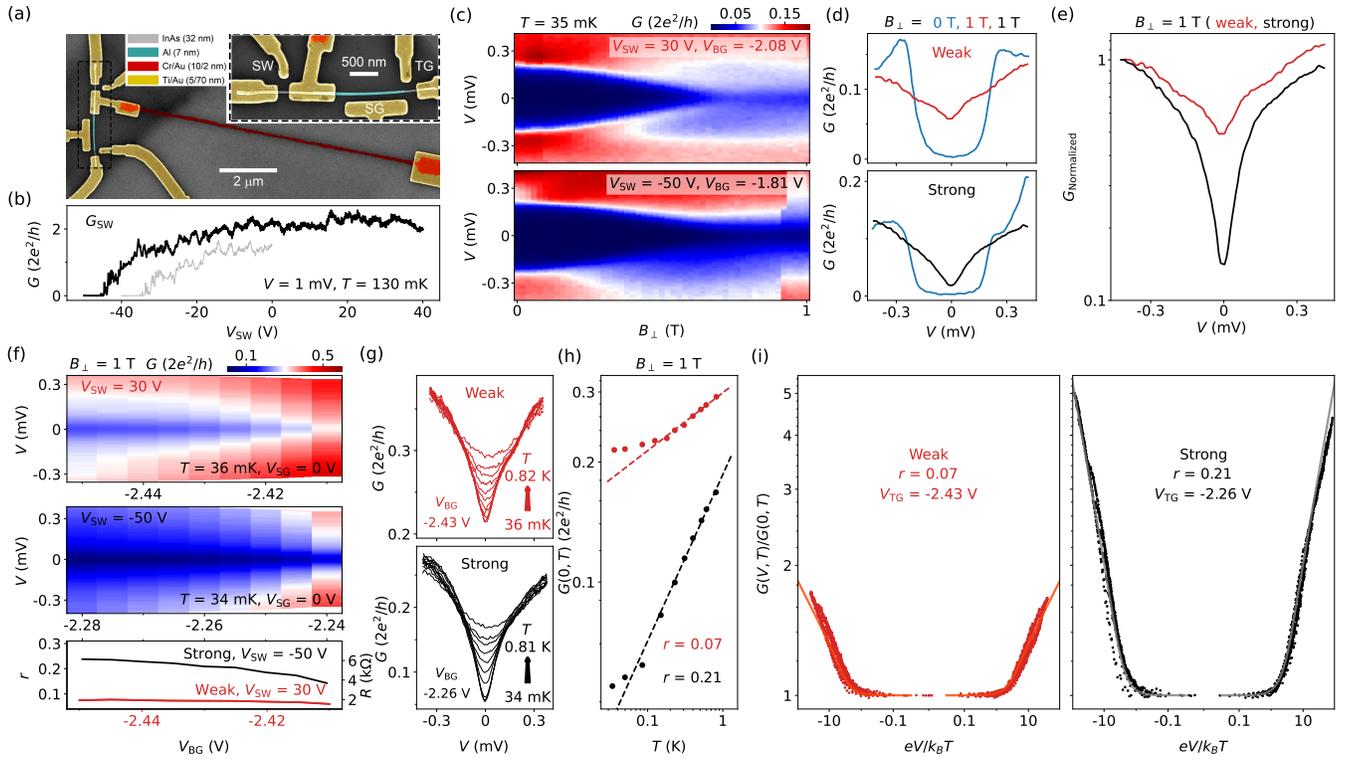

FIG. S4. DCB characterization in device C. All the panels are similar to the ones in Fig. S3.

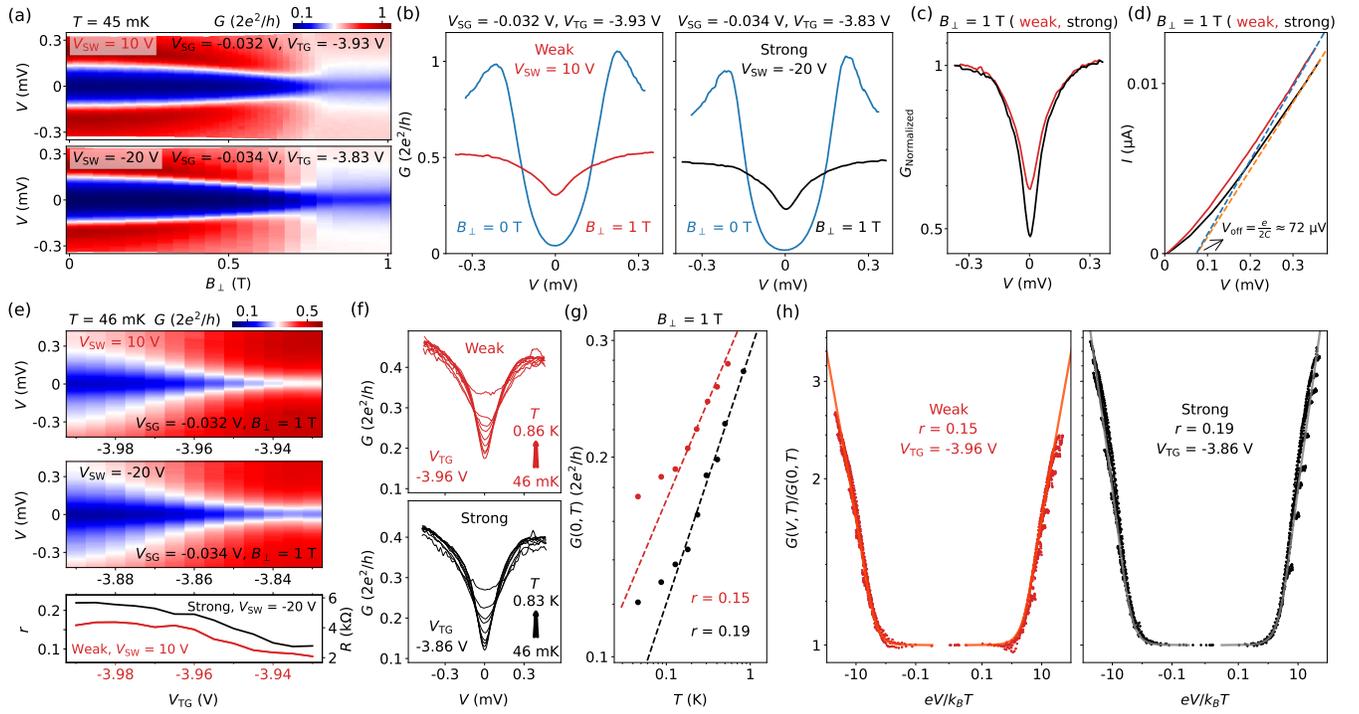

FIG. S5. DCB calibration in device A at high barrier transparencies similar to Figs. 3-4. The red and black curves in (b) are the ones in Figs. 2(e-f). (d) The corresponding $I$-$V$ characteristics of these two curves and the extraction of the charging energy (dashed lines).

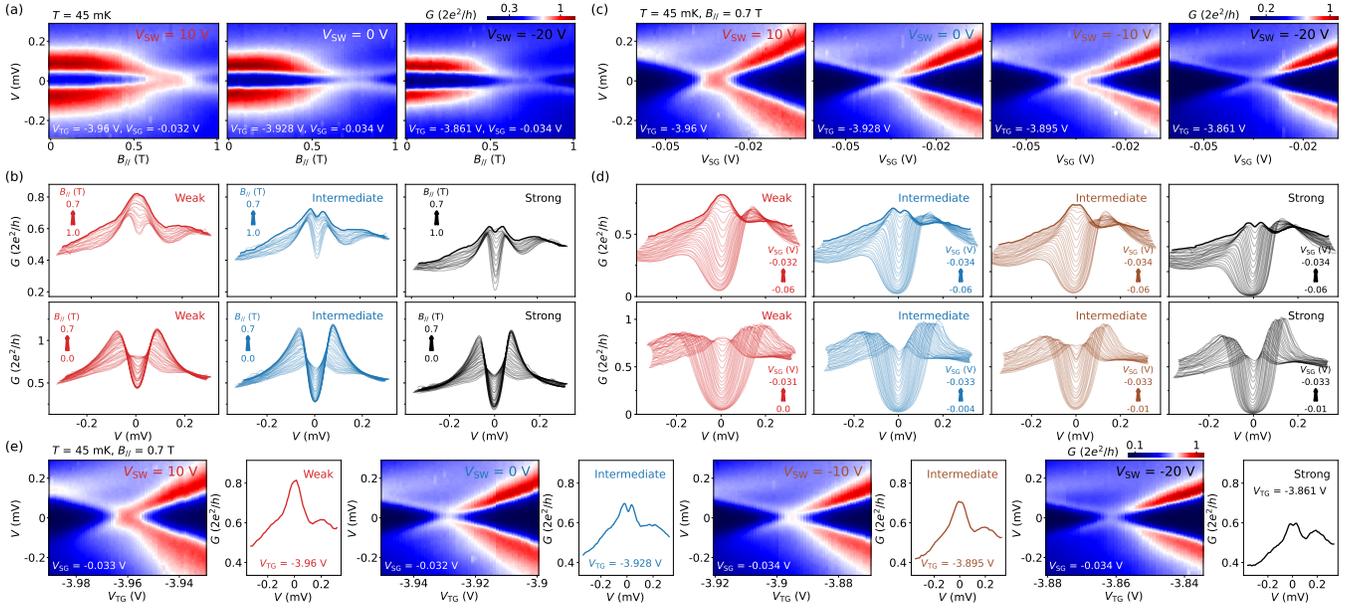

FIG. S6. Additional data of Fig. 3 at intermediate $V_{SW}$ values. (a) $B$ scans of the ZBP at different $V_{SW}$'s. The left and the right ones are the same with Fig. 3(a). (b) Waterfall plots of (a). Lower: 0 to 0.7 T, upper: 0.7 to 1 T. (c) $V_{SG}$ scans of the ZBP at different $V_{SW}$'s. The left and the right ones are the same with Fig. 3(c). (d) Waterfall plots of (c). The case of $V_{SW}$ = -10 V seems to have a stronger DCB strength than the case of $V_{SW}$ = 0 V, suggesting the non-monotonic tuning of DCB by $V_{SW}$, consistent with Fig. S1(b). (e) $V_{TG}$ scans of the ZBP at different $V_{SW}$'s.

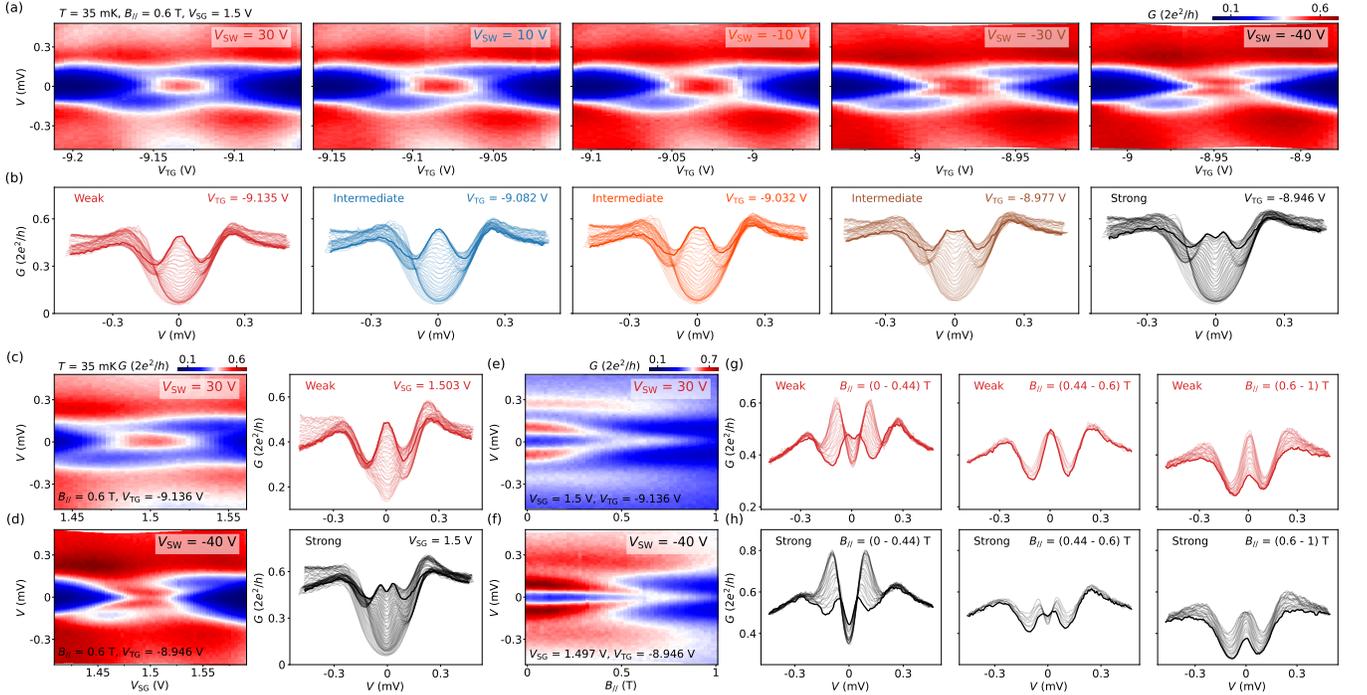

FIG. S7. In situ tuning of DCB on a zero-energy ABS in device B. (a) $V_{TG}$ scans of the ABS at different $V_{SW}$'s. The ZBP in the weak DCB regime (left) becomes split peaks in the strong DCB regime (right). (b) Waterfall plots of (a). (c, d) $V_{SG}$ scans of the ABS in the weak (upper) and strong (lower) DCB regimes. (e, f) $B$ scans of the ABS in the weak (upper) and strong (lower) DCB regimes. (g, h) Line cuts from (e, f). The middle panels show the $B$ range where the ZBP gets split by tuning DCB. The right panels do not show this transition. This is probably due to the small dissipative resistor of device B (Figs. S1(j) and S3(f)): $r$ can only be varied from 0.05 to 0.13.

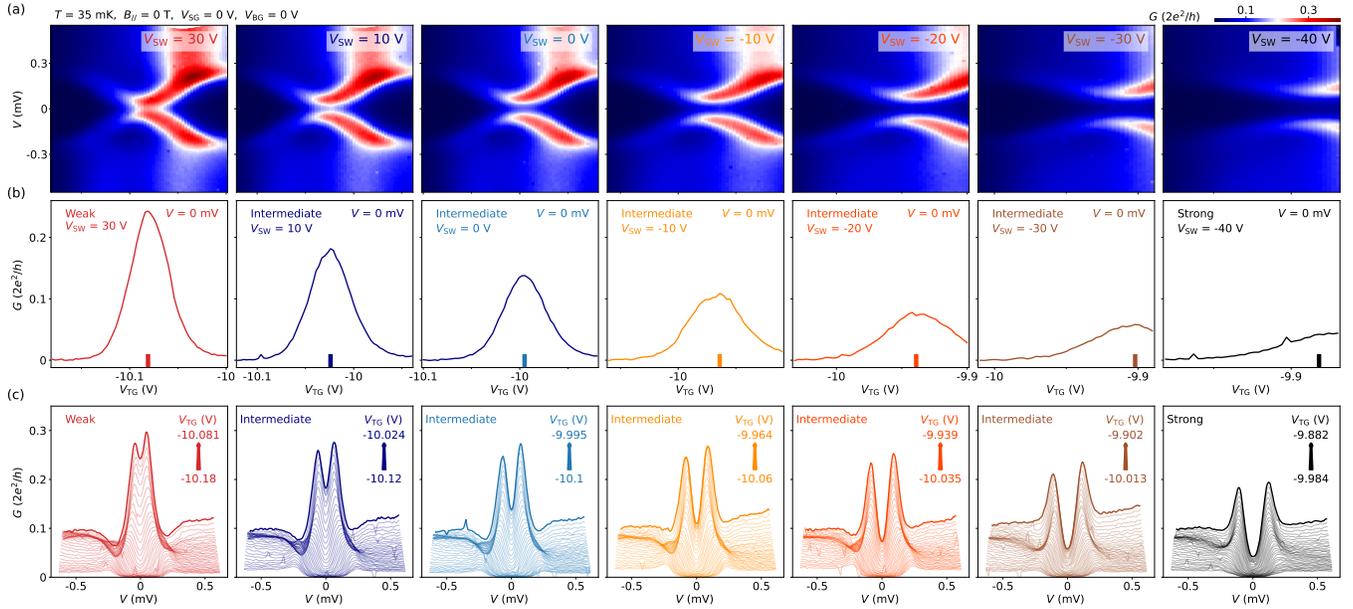

FIG. S8. In situ tuning of DCB on a finite-energy ABS in device B. (a) Gate dependence of a finite-energy ABS (split peaks) at different $V_{SW}$'s. (b) Zero-bias line cuts from (a). (c) Waterfall plots of (a). For clarity, only the gate ranges on the left sides of the bars in (b) are shown. The zero-bias conductance is suppressed more in the strong DCB regime than in the weak DCB regime.

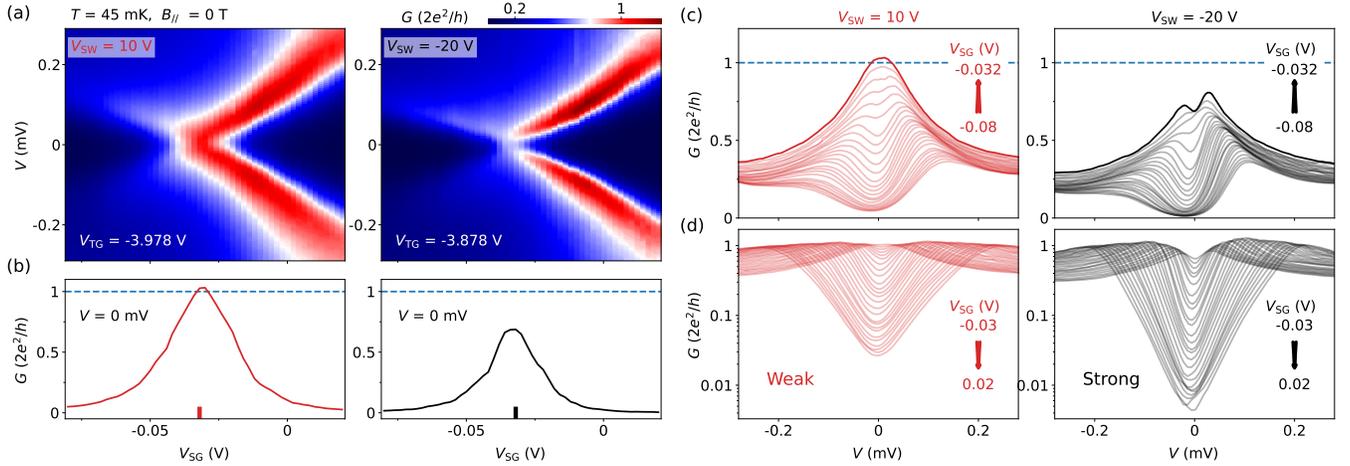

FIG. S9. Additional data for Fig. 4. (a) $V_{SG}$ scans of the ABS. $V_{TG}$ is fixed and indicated by the bars in Fig. 4(b). (b) Zero-bias line cuts from (a). (c) Waterfall plots of (a). The upper panels correspond to the $V_{SG}$ ranges on the left sides of the bars in (b). The lower panels correspond to the $V_{SG}$ ranges on the right sides of the bars in (b). The lower panels use logarithmic scale to highlight the low conductance difference.



\end{document}